\documentclass{emulateapj}
\newcommand{\Fewbody}{{\em Fewbody\/}}

\begin{document}

%%%%%%%%%%%%%%%%%%%%%%%%%%%%%%%%%%%%%%%%%%%%%%%%%%%%%%%%%%%%%%%%%%%%%%%%%%%%%%%
\title{Massive Black Hole Binaries from Collisional Runaways}
\shorttitle{Massive Black Hole Binaries from Collisional Runaways}
\submitted{To appear in \apjl}
\author{M. Atakan G\"urkan\altaffilmark{1}, John M. Fregeau\altaffilmark{2}, and Frederic A. Rasio\altaffilmark{3}}
\shortauthors{G\"urkan, et al.}
\affil{Department of Physics and Astronomy, Northwestern University, Evanston, IL 60208}
\altaffiltext{1}{{\tt ato.gurkan@gmail.com} Current address: Foundations 
Development, Sabanc\i\ University, 34956 \.Istanbul, Turkey.} 
\altaffiltext{2}{\tt fregeau@alum.mit.edu}
\altaffiltext{3}{\tt rasio@northwestern.edu}

%%%%%%%%%%%%%%%%%%%%%%%%%%%%%%%%%%%%%%%%%%%%%%%%%%%%%%%%%%%%%%%%%%%%%%%%%%%%%%%
\begin{abstract}
Recent theoretical work has solidified the viability of the collisional runaway
scenario in young dense star clusters for the formation of very massive stars (VMSs), 
which may be precursors to intermediate-mass black holes (IMBHs).  We present
first results from a numerical study of the collisional runaway process 
in dense star clusters containing primordial binaries.  Stellar collisions during
binary scattering encounters offer an alternate channel for runaway growth,
somewhat independent of direct collisions between single stars.  We find that
clusters with binary fractions $\mathpunct{\gtrsim}10\%$ yield {\em two} VMSs
via collisional runaways, presenting the exotic possibility of forming
IMBH--IMBH binaries in star clusters.  We discuss the implications
for gravitational wave observations, and the impact on cluster
structure.
\end{abstract}

%%%%%%%%%%%%%%%%%%%%%%%%%%%%%%%%%%%%%%%%%%%%%%%%%%%%%%%%%%%%%%%%%%%%%%%%%%%%%%%
\keywords{stellar dynamics --- globular clusters: general --- intermediate-mass black holes 
  --- methods: $n$-body simulations}

%%%%%%%%%%%%%%%%%%%%%%%%%%%%%%%%%%%%%%%%%%%%%%%%%%%%%%%%%%%%%%%%%%%%%%%%%%%%%%%
\section{Introduction}\label{sec:intro}

Observations hinting at the possible existence of intermediate-mass black holes (IMBHs)
have mounted in recent years.  Ultra-luminous X-ray sources---point X-ray
sources with inferred luminosities $\mathpunct{\gtrsim} 10^{39}\,{\rm erg}\,{\rm s}^{-1}$---may
be explained by sub-Eddington accretion onto BHs more massive than the maximum mass
of $\mathpunct{\sim}10M_\sun$ expected via core collapse in main sequence stars,
although viable alternative explanations exist \citep{2004IJMPD..13....1M}.
Similarly, the cuspy core velocity dispersion profiles of the globular clusters
M15 and G1 may also be explained by the dynamical influence of a central IMBH
\citep{2002AJ....124.3255V,2002AJ....124.3270G,2005ApJ...634.1093G}, although 
theoretical work suggests that the observations of M15
may be equally-well explained by a collection of compact stellar remnants in the 
cores of the clusters \citep{2003ApJ...582L..21B}.

At least three distinct IMBH formation mechanisms have been discussed in 
the literature.  The first, and possibly simplest,
is core collapse of a massive Pop III star \citep{2001ApJ...551L..27M}.
The very low metallicity of Pop III stars ($Z\lesssim 10^{-5}Z_\sun$) allows
much larger main-sequence stars to form, limits mass loss during stellar evolution, and increases
the fraction of mass retained in the final BH (for stars more massive than $\mathpunct{\sim}250M_\sun$)
\citep{2001ApJ...550..372F,2004IJMPD..13....1M}.  The second is the successive merging of 
stellar mass BHs via dynamical interactions, which may occur in star clusters that do 
not reach deep core collapse before $\mathpunct{\sim}3\,{\rm Myr}$, when the most massive cluster 
stars have become BHs \citep{1993Natur.364..421K,1993Natur.364..423S,2000ApJ...528L..17P,2002MNRAS.330..232M,oleary2006}.  The process is relatively 
inefficient---in terms of the amount of mass added to the growing BH per BH ejected
from the cluster---requiring BH seeds $\mathpunct{\gtrsim}500 M_\sun$ to create $10^3 M_\sun$ IMBHs 
\citep{2004ApJ...616..221G}, even when aided by the Kozai mechanism \citep{2002MNRAS.330..232M} 
or gravitational wave losses during close approaches \citep{gultekin2005}.
Introducing a mass spectrum for the BHs decreases the required seed mass,
although growth is still rare \citep{oleary2006}.
The third is the runaway merging of main-sequence stars via direct physical
collisions to form a very massive star (VMS), which may then collapse to form an IMBH
\citep{1999A&A...348..117P,2001ApJ...562L..19E,2002ApJ...576..899P,2004ApJ...604..632G}.  
Recent work shows that runaway growth of a VMS occurs generically in clusters with deep 
core collapse times shorter than $\mathpunct{\sim}3\,{\rm Myr}$ \citep{freitag2005b}.

With the exception of one simulation \citep{2004Natur.428..724P}, all
simulations of runaway collisional growth in clusters have ignored the effects of primordial
binaries, which are known to exist in clusters in dynamically significant
numbers \citep{1992PASP..104..981H}.  Indeed, some numerical results suggest that
the primordial binary fraction ($f_b$) may have to be nearly 100\% to explain the currently
observed binary fractions in cluster cores \citep{2005MNRAS.358..572I}.
Primordial binaries are an important piece of the runaway collisional growth
puzzle, since they introduce two effects which may strongly affect the process.
On the one hand, binaries generate energy via dynamical scattering interactions
in cluster cores, supporting the core against deep collapse and limiting
the maximum stellar density attainable, and hence limiting the direct stellar
collision rate \citep{2003gmbp.book.....H}.  On the other hand, stellar
collisions are much more likely in dynamical interactions of binaries, since
the interactions are typically resonant \citep{1996MNRAS.281..830B,2004MNRAS.352....1F}.
Since these two effects of primordial binaries act in opposite
senses with respect to the collision rate, it is not clear {\em a priori} how they
affect the collisional runaway scenario.

Before appealing to numerical methods, however, one can gain insight
into the effects of primordial binaries by considering the coagulation
equation \citep{1993ApJ...418..147L,2000Icar..143...74L,2001Icar..150..314M}---a 
simplification of which is presented in \citet{freitag2005a}---which describes
the evolution of a spectrum of masses due to mergers.  For growth to occur in a
runaway fashion, the coagulation equation requires that the cross section for 
collisions with the 
runaway object scales sufficiently rapidly with its mass: $S_{\rm coll} \propto M^\eta$, with
$\eta > 1$.  For single--single star collisions in star clusters in which the central
velocity dispersion is less than the escape speed from the surface of a typical star (so
that the cross section is dominated by gravitational focusing), this corresponds to 
the constraint $R \propto M^\alpha$ with $\alpha > 0$ on the main-sequence mass--radius 
relationship, which is satisfied by main-sequence stars of any mass or metallicity.
With some approximations, the coagulation equation analysis \citep{1975MNRAS.173..729H,1983ApJ...268..319H,1993ApJ...415..631S,gultekin2005} can be applied to collisions
occurring in binary scattering interactions.  From Fig.~4 of \citet{gultekin2005},
the cross section for close approach distances of $r_{\rm min}$ in binary--single
scattering encounters scales as $(S_{\rm coll}/\pi a^2) (v_\infty/v_c)^2 \propto (r_{\rm min}/a)^\gamma$, 
with $0.3 \lesssim \gamma \lesssim 1$, where $a$ is the binary semimajor axis, $v_\infty$
is the relative velocity between the binary and single star at infinity, and $v_c$ is the
critical velocity \citep[see {eq.~[12]} of][]{gultekin2005}.  Using the radius of the
runaway star for $r_{\rm min}$, $R \propto M^\beta$ with 
$0.5 \lesssim \beta \lesssim 1$ for the scaling of the mass--radius relation, and 
assuming that the binding energy of the binary is roughly preserved during the encounter
so that $a \propto M$, we find: 
\begin{equation}\label{eq:coag}
 S_{\rm coll} \propto M^{2 + \gamma (\beta - 1)}\,,
\end{equation}
with $0.3 \lesssim \gamma \lesssim 1$ and $0.5 \lesssim \beta \lesssim 1$.  The 
minimum of the exponent is $\mathpunct{\approx} 1.5$.  Thus according to the coagulation equation,
collisions induced in binary--single scattering interactions should yield runaway
growth of a VMS.  This result says nothing about the {\em rate} of growth of
the runaway object.  In other words, it is still not clear whether binary interactions
will limit the cluster core density such that the runaway timescale is longer
than the massive star main-sequence lifetime of $\mathpunct{\approx} 3\,{\rm Myr}$, in which
case the process would be halted.

Assuming that the cluster core density
reached is high enough for the runaway to proceed, it would appear that a
single binary is sufficient for a binary interaction-induced runaway to occur.
This is, of course, not the case, since binary scattering interactions
tend to destroy binaries.  Thus we expect that for sufficiently low $f_b$,
the runaway will be primarily mediated by single--single collisions.  For
sufficiently large $f_b$, the runaway will be primarily mediated by binary--binary
interactions (the analysis above assumes binary--single interactions---for
binary--binary the value of $\gamma$ will be smaller, still allowing a 
runaway by eq.~[\ref{eq:coag}]).  For intermediate $f_b$ it is possible that a binary interaction
induced-runaway could proceed until the core binary population is sufficiently
depleted that the cluster's core collapses.  If the first runaway is 
far enough from the center of the cluster when the core collapses, it 
is possible that a second runaway will be formed during core collapse,
mediated by single--single collisions.  The exotic possibility of forming 
two VMSs in a cluster, and thus two IMBHs, is a tantalizing one, with 
implications for gravitational wave observations and cluster dynamics.

In this letter we present first results from a study of the runaway collisional
scenario for the formation of VMSs in young dense clusters with primordial binaries.
We briefly describe our numerical method, present results showing the growth of
two runaways, and discuss the implications.

%%%%%%%%%%%%%%%%%%%%%%%%%%%%%%%%%%%%%%%%%%%%%%%%%%%%%%%%%%%%%%%%%%%%%%%%%%%%%%%
\section{Numerical Analysis}\label{sec:analysis}

%\subsection{Method and Initial Conditions}\label{subsec:method}

We use our Monte Carlo cluster code to simulate the evolution
of young star clusters with primordial binaries 
\citep{2000ApJ...540..969J,2003ApJ...593..772F,2004ApJ...604..632G}.
This code uses the H\'enon method for two-body relaxation,
and incorporates the \Fewbody\ $N$-body integrator to perform
dynamical scattering encounters of binaries.  Stellar collisions are handled
by assuming that stars whose surfaces touch merge with
no mass loss, an assumption that has been shown to be valid
for clusters with low velocity dispersion, such as globulars or
young dense clusters \citep{freitag2005a,freitag2005b}.  
Collisions are allowed to occur directly in single--single star encounters, 
and during binary interactions \citep{2004MNRAS.352....1F}.

\begin{figure}
  \begin{center}
    \includegraphics[width=\columnwidth]{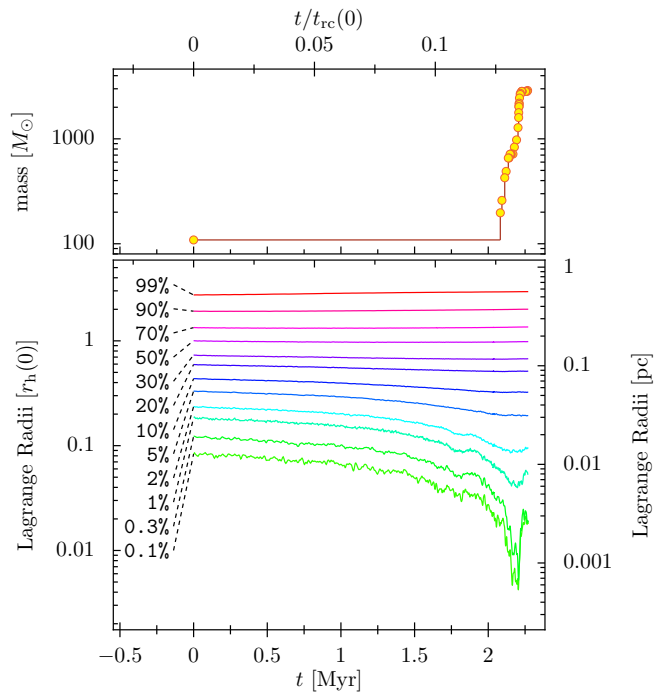}
      \caption{Evolution of the cluster Lagrange radii (lower panel) in units of the 
	initial cluster half-mass radius (left axis) and in parsecs (right axis),
	and mass of the (single) runaway for a cluster with $10^6$ objects and 
	$f_b=0.05$.  Time is shown both in Myr and in units of the central
	relaxation time.\label{fig:lr605}}
  \end{center}
\end{figure}

For initial conditions we assume a $W_0=3$ King model density
profile for the cluster, with no initial mass segregation.  
All stars are main-sequence stars with masses in the range 
$0.2<M/M_\odot<120$, distributed
according to a Salpeter mass function.  Simulations show that
the primary condition required for a runaway is that
the core collapse time is shorter than the main-sequence
lifetime for the most massive stars, $\mathpunct{\sim}3\,{\rm Myr}$ \citep{freitag2005b};
and that for clusters of single stars
with a wide mass spectrum, the core collapse time is
always 
%\begin{equation}\label{eq:tcc}
$t_{\rm cc} \approx 0.15\, t_{\rm rc} (0)$,
%\end{equation}
(where $t_{\rm rc}(0)$ is the initial relaxation time in the core),
independent of cluster mass, size, or density profile
\citep{2004ApJ...604..632G}.  This allows us to set the central density
of the cluster such that the predicted core collapse time 
is either less or greater than 
$3\,{\rm Myr}$.  We perform simulations with binary 
fractions\footnote{The binary fraction is defined
such that $f_b$ is the fraction of {\em objects} in the cluster
that are binaries, an object being either a binary or a single
star.} up to $0.2$.  The binary population is created from a cluster
of single stars by adding secondary companions to randomly chosen
cluster stars, with the secondary mass chosen uniformly in the 
binary mass ratio.  The binary binding energy is distributed
uniformly in the logarithm, truncated at high energy so that
the binary members do not make contact at pericenter, and truncated
at low energy so that the orbital speed of the lightest member
in the binary is larger than the local stellar velocity dispersion
\citep[see][]{fregeau2005}.  The eccentricity is chosen according
to a thermal distribution, truncated at large $e$ so that the binary
members do not make contact at pericenter.  We use either
$N=5\times 10^5$ or $10^6$ total cluster objects, finding no
difference in the runaway results between the two.  We run all simulations
until $3\,{\rm Myr}$.  Our criterion for a runaway is that its final
mass is $\mathpunct{\gtrsim}500 M_\sun$.  In our runs the final masses of the VMSs 
are always at least twice this value.

%%%%%%%%%%%%%%%%%%%%%%%%%%%%%%%%%%%%%%%%%%%%%%%%%%%%%%%%%%%%%%%%%%%%%%%%%%%%%%%
%\subsection{Results}\label{subsec:results}
\begin{figure}
  \begin{center}
    \includegraphics[width=\columnwidth]{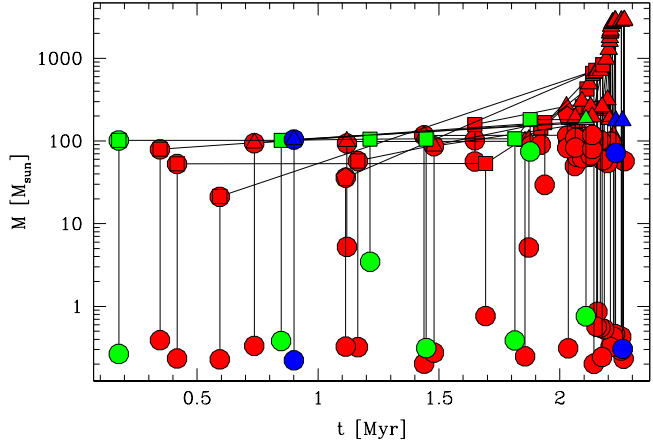}
    \caption{Merger trees for the three most massive stars at the end of the simulation
      for the model presented in Figure~\ref{fig:lr605}.  Red shows the most massive,
      intermediate green the second most, and blue the third most.  For each collision with the massive
      star, a symbol is plotted at its mass at the time of the 
      collision with lines drawn connecting to the symbols representing the star 
      participating in the merger.  A triangle represents a single--single collision, a 
      square a collision in a binary--single interaction, a pentagon a collision in
      a binary--binary interaction, and a circle a star that has not yet undergone a 
      collision.  As an example, the three intermediate green points at $t\approx 0.2\,{\rm Myr}$
      represent a collision in a binary--single interaction of a $\mathpunct{\approx} 100 M_\sun$
      star with a $\mathpunct{\approx} 0.3 M_\sun$ star, each having never undergone a collision
      previously.  For this model there is clearly just one VMS, created 
      primarily in single--single collisions.\label{fig:mt605}}
  \end{center}
  \vspace{-0.5cm}
\end{figure}

Figure~\ref{fig:lr605} shows the evolution of the cluster Lagrange
radii and the mass of the (single) runaway as a function of time
for a cluster with $t_{\rm cc} < 3\,{\rm Myr}$ %(as predicted by
%eq.~[\ref{eq:tcc}]) 
and $f_b=0.05$.
The evolution is typical for models with $f_b \lesssim 0.1$
and $t_{\rm cc} < 3\,{\rm Myr}$ in that: 1) the energy generated 
in binary interactions is not sufficient to postpone core collapse 
beyond $3\,{\rm Myr}$, and 2) there is a single runaway.
In the model shown in Figure~\ref{fig:lr605}, the runaway
grows to $\mathpunct{\approx} 2800 M_\sun$ before the cluster begins
to expand in response to the energy generated in collisions
with the runaway.  

\begin{figure}
  \begin{center}
    \includegraphics[width=\columnwidth]{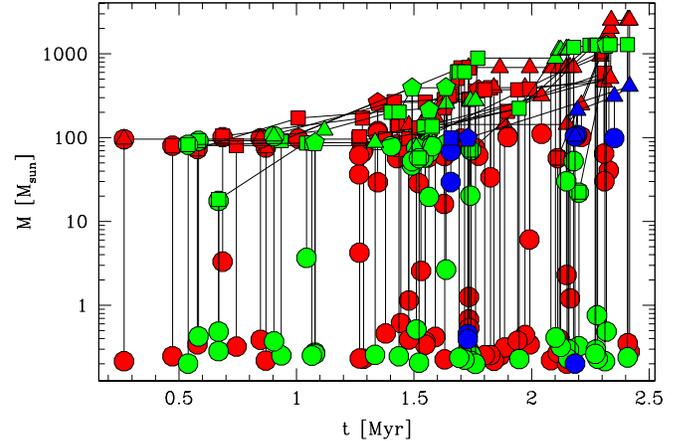}
    \caption{Same as Figure~\ref{fig:mt605} for the same model but with
      $f_b=0.1$.  A binary interaction-induced collisional runaway begins at 
      $t\approx 1.5\,{\rm Myr}$ and proceeds to $\mathpunct{\approx} 1.3 \times 10^3\,M_\sun$.  
      A direct collision-induced runaway begins at $t\approx 2.3\,{\rm Myr}$, yielding
      a second runaway, of mass $\mathpunct{\approx} 2.5\times 10^3\,M_\sun$.  The third most 
      massive star at the end of the simulation has only grown to $\mathpunct{\approx} 400 M_\sun$.
      \label{fig:mt610}}
  \end{center}
  \vspace{-0.5cm}
\end{figure}

We follow the membership of the collision products in binaries during
the collisions and binary interactions. This allows us to produce
merger trees to track the contributions to the formation of VMSs.
Figure~\ref{fig:mt605} shows the merger 
trees for the three most massive stars at the end of the simulation.
There is clearly only one runaway for this model, and it grows
primarily via direct, single--single collisions.
Figure~\ref{fig:mt610} shows the same as Figure~\ref{fig:mt605}
for the same model but with $f_b=0.1$.  In this model there are
two VMSs at the end of the simulation.  A binary interaction-induced 
collisional runaway begins at 
$t\approx 1.5\,{\rm Myr}$ and proceeds to $\mathpunct{\approx} 1.3 \times 10^3\,M_\sun$.  
A direct collision-induced runaway begins at $t\approx 2.3\,{\rm Myr}$, yielding
a second runaway, of mass $\mathpunct{\approx} 2.5\times 10^3\,M_\sun$.  The third most 
massive star at the end of the simulation has only grown to $\mathpunct{\approx} 400 M_\sun$.
The evolution is typical for models with $f_b \gtrsim 0.1$
and $t_{\rm cc} < 3\,{\rm Myr}$ in that there are two runaways: 
one via direct collisions, and one via collisions in binary interactions.

Although we have not yet conducted a full parameter space survey,
we have mapped a significant slice in $f_b$--$t_{\rm rc}(0)$ space,
from $f_b=0.02$ to $0.2$ and from $t_{\rm cc}/(3\,{\rm Myr})=0.16$ to $28$.
Without exception, and independent of $f_b$, models with 
$t_{\rm cc}=0.15\,t_{\rm rc}(0) > 3\,{\rm Myr}$ show no runaways (7 models), while
those with $t_{\rm cc}=0.15\,t_{\rm rc}(0) < 3\,{\rm Myr}$ show either one or two (10 models).
This is in agreement with the results for clusters with no primordial
binaries, presumably since binary fractions smaller than 20\% are not
sufficient to postpone core collapse beyond $3\,{\rm Myr}$.
Of the models that show runaways, those with $f_b < 0.1$ always yield
only single runaways (4 models), while those with $f_b \geq 0.1$ always yield 
double runaways (6 models).  No models ever produce more than two 
runaways.  Due to the computational cost involved, we have not yet explored 
the $f_b>0.2$ region of parameter space.  

The numerical results presented here clearly agree with the
analytical arguments made above on the runaway nature of binary
interaction-induced collisions. Binary interaction-induced
runaways occur in the outskirts or just outside the core (typically
a few percents of a parsec for our models). This is a region where
densities, in particular the binary densities, are high enough to
start a runaway but the relaxation time is long enough that massive
objects do not rapidly sink to the center.  The rapid collapse near
the center leads to an expansion of these regions, further increasing
the relaxation time and prevents a prompt merger of the VMSs. 

Finally, we note the apparent disagreement between our results and those of 
\citet{2004Natur.428..724P}, who performed two direct $N$-body integrations
of models of the cluster MGG-11 with $f_b=0.1$ and found only single runaways.  
Our results predict that whenever $f_b \gtrsim 0.1$, and the cluster
central relaxation time is short enough so that a runaway is expected
in a corresponding cluster with only single stars, a double runaway
should result.  The models of \citet{2004Natur.428..724P} are
extremely centrally concentrated, with $W_0=12$, while our models
all use $W_0=3$.  From the discussion above, it is clear that if the 
density profile is sufficiently steep, the binary interaction-induced runaway will 
take place close enough to the center so that it will seed the subsequent
direct collision-induced runaway, yielding just one runaway.  
In addition, the condition $f_b \gtrsim 0.1$
for double runaways is only approximate.  Since the models of 
\citet{2004Natur.428..724P} are right on this boundary, it is possible
that a small difference in the initial conditions has prevented a double
runaway in their models.

%%%%%%%%%%%%%%%%%%%%%%%%%%%%%%%%%%%%%%%%%%%%%%%%%%%%%%%%%%%%%%%%%%%%%%%%%%%%%%%
\section{Discussion}\label{sec:discussion}

Although the process is somewhat uncertain, it is likely that 
the VMSs formed in young clusters via collisional runaways 
will undergo core-collapse supernovae
and become IMBHs on a timescale of $\mathpunct{\sim}1\,{\rm Myr}$ \citep{freitag2005b}.  
Our results show no evidence of 
the VMSs merging prior to becoming IMBHs.  After their separate formation,
the two IMBHs will quickly exchange into a common binary via dynamical
interactions.  The IMBH--IMBH binary will then shrink via dynamical friction
due to the cluster stars, to the point at which the stellar mass enclosed
in the binary is comparable to the binary mass.  This occurs on a timescale
$\mathpunct{\sim}t_r \langle m \rangle / M_{\rm IMBH}$, where $t_r$ is the
local relaxation time and $\langle m \rangle$ is the local average stellar
mass.  Since $\langle m \rangle / M_{\rm IMBH} \lesssim 10^{-2}$, this timescale
is likely to be $\mathpunct{\lesssim}10\,{\rm Myr}$.  The binary will then shrink
via dynamical encounters with cluster stars, the rate of which
is governed by loss-cone physics.  The timescale for the binary to shrink
to the point at which it merges quickly via gravitational
radiation energy loss is likely to be $\mathpunct{\lesssim}1\,{\rm Gyr}$ 
\citep{2003ApJ...599.1129Y,2005ApJ...618..426M}.  

Although IMBH--IMBH binaries do not merge in the LISA band---the gravitational wave frequency at
merger is $\mathpunct{\sim}1\,{\rm Hz}$---they do represent bright
sources that take at least $\mathpunct{\sim}10^6\,{\rm yr}$ to cross the LISA band.  
Their inspiral (chirp) signals
should be easily detectable by LISA out to a few tens of Mpc. Thus the
number of detectable IMBH binary sources may be quite large, since
most clusters are probably born with $f_b \gtrsim 0.1$, and
any cluster with mass $\mathpunct{\gtrsim}10^6 M_\sun$ and central relaxation time 
$\mathpunct{\lesssim}20\,{\rm Myr}$ will lead to a double runaway.

Significant core rotation (with rotational speed comparable to the local
velocity dispersion) is observed in the clusters M15, $\omega$ Cen, 47 Tuc, and G1
\citep[e.g.][]{2002AJ....124.3255V,2005ApJ...634.1093G}.
This rotation suggests the presence of a core angular momentum source, 
such as an IMBH binary \citep{2005MNRAS.364.1315M}.  
Similarly, observations of a millisecond pulsar in the halo of NGC 6752, and two
others in the core with high negative spin derivatives, hint at the existence
of an IMBH binary in the core \citep{2003ApJ...599.1260C}.
Since the IMBH--IMBH binaries formed via collisional runaways will merge 
within $\mathpunct{\sim}1\,{\rm Gyr}$ after formation, any angular momentum imparted to 
the cluster by the IMBH--IMBH binary will quickly diffuse out of the core on a 
core relaxation time.  An alternate mechanism must be at work in creating
the core rotation seen in some globular clusters today.

%%%%%%%%%%%%%%%%%%%%%%%%%%%%%%%%%%%%%%%%%%%%%%%%%%%%%%%%%%%%%%%%%%%%%%%%%%%%%%%
\acknowledgements
\vspace{-0.6cm}
We thank Marc Freitag, Shane Larson, and Cole Miller for many fruitful
discussions, and an anonymous referee for comments which improved the paper.
Some of the
simulations were performed on the Tungsten cluster at the National Center for 
Supercomputing Applications.  MAG thanks Sabanc\i{} University 
for their hospitality during the finalization of this letter.
This work was supported by NASA grant NNG04G176G.

%%%%%%%%%%%%%%%%%%%%%%%%%%%%%%%%%%%%%%%%%%%%%%%%%%%%%%%%%%%%%%%%%%%%%%%%%%%%%%%
\bibliographystyle{apj}

\begin{thebibliography}{33}
\expandafter\ifx\csname natexlab\endcsname\relax\def\natexlab#1{#1}\fi

\bibitem[{{Bacon} {et~al.}(1996){Bacon}, {Sigurdsson}, \&
  {Davies}}]{1996MNRAS.281..830B}
{Bacon}, D., {Sigurdsson}, S., \& {Davies}, M.~B. 1996, \mnras, 281, 830

\bibitem[{{Baumgardt} {et~al.}(2003){Baumgardt}, {Hut}, {Makino}, {McMillan},
  \& {Portegies Zwart}}]{2003ApJ...582L..21B}
{Baumgardt}, H., {Hut}, P., {Makino}, J., {McMillan}, S., \& {Portegies Zwart},
  S. 2003, \apjl, 582, L21

\bibitem[{{Colpi} {et~al.}(2003){Colpi}, {Mapelli}, \&
  {Possenti}}]{2003ApJ...599.1260C}
{Colpi}, M., {Mapelli}, M., \& {Possenti}, A. 2003, \apj, 599, 1260

\bibitem[{{Ebisuzaki} {et~al.}(2001){Ebisuzaki}, {Makino}, {Tsuru}, {Funato},
  {Portegies Zwart}, {Hut}, {McMillan}, {Matsushita}, {Matsumoto}, \&
  {Kawabe}}]{2001ApJ...562L..19E}
{Ebisuzaki}, T., {Makino}, J., {Tsuru}, T.~G., {Funato}, Y., {Portegies Zwart},
  S., {Hut}, P., {McMillan}, S., {Matsushita}, S., {Matsumoto}, H., \&
  {Kawabe}, R. 2001, \apjl, 562, L19

\bibitem[{{Fregeau} {et~al.}(2005){Fregeau}, {Chatterjee}, \&
  {Rasio}}]{fregeau2005}
{Fregeau}, J.~M., {Chatterjee}, S., \& {Rasio}, F.~A. 2005, \apj, in press
  (astro-ph/0510748)

\bibitem[{{Fregeau} {et~al.}(2004){Fregeau}, {Cheung}, {Portegies Zwart}, \&
  {Rasio}}]{2004MNRAS.352....1F}
{Fregeau}, J.~M., {Cheung}, P., {Portegies Zwart}, S.~F., \& {Rasio}, F.~A.
  2004, \mnras, 352, 1

\bibitem[{{Fregeau} {et~al.}(2003){Fregeau}, {G{\"u}rkan}, {Joshi}, \&
  {Rasio}}]{2003ApJ...593..772F}
{Fregeau}, J.~M., {G{\"u}rkan}, M.~A., {Joshi}, K.~J., \& {Rasio}, F.~A. 2003,
  \apj, 593, 772

\bibitem[{{Freitag} {et~al.}(2005{\natexlab{a}}){Freitag}, {G\"urkan}, \&
  {Rasio}}]{freitag2005b}
{Freitag}, M., {G\"urkan}, M.~A., \& {Rasio}, F.~A. 2005{\natexlab{a}}, \mnras,
  in press (astro-ph/0503130)

\bibitem[{{Freitag} {et~al.}(2005{\natexlab{b}}){Freitag}, {Rasio}, \&
  {Baumgardt}}]{freitag2005a}
{Freitag}, M., {Rasio}, F.~A., \& {Baumgardt}, H. 2005{\natexlab{b}}, \mnras,
  in press (astro-ph/0503129)

\bibitem[{{Fryer} {et~al.}(2001){Fryer}, {Woosley}, \&
  {Heger}}]{2001ApJ...550..372F}
{Fryer}, C.~L., {Woosley}, S.~E., \& {Heger}, A. 2001, \apj, 550, 372

\bibitem[{{Gebhardt} {et~al.}(2005){Gebhardt}, {Rich}, \&
  {Ho}}]{2005ApJ...634.1093G}
{Gebhardt}, K., {Rich}, R.~M., \& {Ho}, L.~C. 2005, \apj, 634, 1093

\bibitem[{{Gerssen} {et~al.}(2002){Gerssen}, {van der Marel}, {Gebhardt},
  {Guhathakurta}, {Peterson}, \& {Pryor}}]{2002AJ....124.3270G}
{Gerssen}, J., {van der Marel}, R.~P., {Gebhardt}, K., {Guhathakurta}, P.,
  {Peterson}, R.~C., \& {Pryor}, C. 2002, \aj, 124, 3270

\bibitem[{{G{\"u}ltekin} {et~al.}(2004){G{\"u}ltekin}, {Miller}, \&
  {Hamilton}}]{2004ApJ...616..221G}
{G{\"u}ltekin}, K., {Miller}, M.~C., \& {Hamilton}, D.~P. 2004, \apj, 616, 221

\bibitem[{{G\"ultekin} {et~al.}(2005){G\"ultekin}, {Miller}, \&
  {Hamilton}}]{gultekin2005}
{G\"ultekin}, K., {Miller}, M.~C., \& {Hamilton}, D.~P. 2005, \apj, in press
  (astro-ph/0509885)

\bibitem[{{G{\"u}rkan} {et~al.}(2004){G{\"u}rkan}, {Freitag}, \&
  {Rasio}}]{2004ApJ...604..632G}
{G{\"u}rkan}, M.~A., {Freitag}, M., \& {Rasio}, F.~A. 2004, \apj, 604, 632

\bibitem[Heggie(1975)]{1975MNRAS.173..729H} Heggie, D.~C.\ 1975, \mnras, 
173, 729

\bibitem[{{Heggie} \& {Hut}(2003)}]{2003gmbp.book.....H}
{Heggie}, D. \& {Hut}, P. 2003, {The Gravitational Million-Body Problem} (Cambridge University Press)

\bibitem[{{Hut} {et~al.}(1992){Hut}, {McMillan}, {Goodman}, {Mateo}, {Phinney},
  {Pryor}, {Richer}, {Verbunt}, \& {Weinberg}}]{1992PASP..104..981H}
{Hut}, P., {McMillan}, S., {Goodman}, J., {Mateo}, M., {Phinney}, E.~S.,
  {Pryor}, C., {Richer}, H.~B., {Verbunt}, F., \& {Weinberg}, M. 1992, \pasp,
  104, 981

\bibitem[Hut \& Bahcall(1983)]{1983ApJ...268..319H} Hut, P., \& Bahcall, 
J.~N.\ 1983, \apj, 268, 319 

\bibitem[{{Ivanova} {et~al.}(2005){Ivanova}, {Belczynski}, {Fregeau}, \&
  {Rasio}}]{2005MNRAS.358..572I}
{Ivanova}, N., {Belczynski}, K., {Fregeau}, J.~M., \& {Rasio}, F.~A. 2005,
  \mnras, 358, 572

\bibitem[{{Joshi} {et~al.}(2000){Joshi}, {Rasio}, \& {Portegies
  Zwart}}]{2000ApJ...540..969J}\looseness-1
{Joshi}, K.~J., {Rasio}, F.~A., \& {Portegies Zwart}, S. 2000, \apj, 540, 969

\bibitem[Kulkarni et al.(1993)]{1993Natur.364..421K} Kulkarni, S.~R., Hut, 
P., \& McMillan, S.\ 1993, \nat, 364, 421 

\bibitem[{{Lee}(1993)}]{1993ApJ...418..147L}
{Lee}, M.~H. 1993, \apj, 418, 147

\bibitem[{{Lee}(2000)}]{2000Icar..143...74L}
---. 2000, Icarus, 143, 74

\bibitem[{{Madau} \& {Rees}(2001)}]{2001ApJ...551L..27M}
{Madau}, P. \& {Rees}, M.~J. 2001, \apjl, 551, L27

\bibitem[{{Malyshkin} \& {Goodman}(2001)}]{2001Icar..150..314M}
{Malyshkin}, L. \& {Goodman}, J. 2001, Icarus, 150, 314

\bibitem[{{Mapelli} {et~al.}(2005){Mapelli}, {Colpi}, {Possenti}, \&
  {Sigurdsson}}]{2005MNRAS.364.1315M}
{Mapelli}, M., {Colpi}, M., {Possenti}, A., \& {Sigurdsson}, S. 2005, \mnras,
  364, 1315

\bibitem[{{Miller}(2005)}]{2005ApJ...618..426M}
{Miller}, M.~C. 2005, \apj, 618, 426

\bibitem[{{Miller} \& {Colbert}(2004)}]{2004IJMPD..13....1M}
{Miller}, M.~C. \& {Colbert}, E.~J.~M. 2004, Int. J. Mod.
  Phy. D, 13, 1

\bibitem[{{Miller} \& {Hamilton}(2002)}]{2002MNRAS.330..232M}
{Miller}, M.~C. \& {Hamilton}, D.~P. 2002, \mnras, 330, 232

%\bibitem[{{O'Leary} {et~al.}(2005){O'Leary}, {Rasio}, {Fregeau}, {Ivanova}, \&
%  {O'Shaughnessy}}]{oleary2005}
%{O'Leary}, R.~M., {Rasio}, F.~A., {Fregeau}, J.~M., {Ivanova}, N., \&
%  {O'Shaughnessy}, R. 2005, \apj, accepted (astro-ph/0508224)
\bibitem[O'Leary et al.(2006)]{oleary2006} O'Leary, R.~M., Rasio, 
F.~A., Fregeau, J.~M., Ivanova, N., \& O'Shaughnessy, R.\ 2006, \apj, 637, 
937 

\bibitem[{{Portegies Zwart} {et~al.}(2004){Portegies Zwart}, {Baumgardt},
  {Hut}, {Makino}, \& {McMillan}}]{2004Natur.428..724P}
{Portegies Zwart}, S.~F., {Baumgardt}, H., {Hut}, P., {Makino}, J., \&
  {McMillan}, S.~L.~W. 2004, \nat, 428, 724

\bibitem[{{Portegies Zwart} {et~al.}(1999){Portegies Zwart}, {Makino},
  {McMillan}, \& {Hut}}]{1999A&A...348..117P}
{Portegies Zwart}, S.~F., {Makino}, J., {McMillan}, S.~L.~W., \& {Hut}, P.
  1999, \aap, 348, 117

\bibitem[Portegies Zwart \& McMillan(2000)]{2000ApJ...528L..17P} Portegies 
Zwart, S.~F., \& McMillan, S.~L.~W.\ 2000, \apjl, 528, L17 

\bibitem[{{Portegies Zwart} \& {McMillan}(2002)}]{2002ApJ...576..899P}
{Portegies Zwart}, S.~F. \& {McMillan}, S.~L.~W. 2002, \apj, 576, 899

\bibitem[Sigurdsson \& Hernquist(1993)]{1993Natur.364..423S} Sigurdsson, 
S., \& Hernquist, L.\ 1993, \nat, 364, 423

\bibitem[Sigurdsson \& Phinney(1993)]{1993ApJ...415..631S} Sigurdsson, S., 
\& Phinney, E.~S.\ 1993, \apj, 415, 631 

\bibitem[{{van der Marel} {et~al.}(2002){van der Marel}, {Gerssen},
  {Guhathakurta}, {Peterson}, \& {Gebhardt}}]{2002AJ....124.3255V}
{van der Marel}, R.~P., {Gerssen}, J., {Guhathakurta}, P., {Peterson}, R.~C.,
  \& {Gebhardt}, K. 2002, \aj, 124, 3255

\bibitem[{{Yu} \& {Tremaine}(2003)}]{2003ApJ...599.1129Y}
{Yu}, Q. \& {Tremaine}, S. 2003, \apj, 599, 1129

\end{thebibliography}

\clearpage

\clearpage

%%%%%%%%%%%%%%%%%%%%%%%%%%%%%%%%%%%%%%%%%%%%%%%%%%%%%%%%%%%%%%%%%%%%%%%%%%%%%%%

\end{document}